# Active Correlation Technique: Status and Development

Yury Tsyganov

*FLNR, JINR. 141980 Dubna, Moscow Region, Joliot-Curie str.6*

**Abstract.** During the recent years, at the FLNR (JINR) a successful cycle of experiments has been accomplished on the synthesis of the superheavy elements with Z=112-118 with $^{48}$Ca beam. From the viewpoint of the detection of rare decays and background suppression, this success was achieved due to the application of a new radical technique – the method of active correlations. The method employs search in a real-time mode for a pointer to a probable correlation like recoil-alpha for switching the beam off. In the case of detection in the same detector strip an additional alpha-decay event, of "beam OFF" time interval is prolonged automatically

## INTRODUCTION

The During the recent years, at the FLNR (JINR) a successful cycle of experiments has been accomplished on the synthesis of the super heavy elements with Z=112-118 with $^{48}$Ca beam [1-7]. Usually, to reach high total SHE experiment efficiency, one use extremely high ($n*10^{12}$ to $10^{13}$ pps, n > 1) heavy ion beam intensities. It means, that not only irradiated target, sometimes (frequently) made on highly radioactive actinide material, should not be destroyed during long term experiment, but the in-flight recoil separator and its detection system should provide backgrounds suppression in order to extract one-two events from the whole data flow. Typically, the DGFRS provides suppression of the beam-like and target-like backgrounds by the factors of[1] ~$10^{15}$-$10^{17}$ and $10^4$-$5*10^4$, respectively . Nevertheless, under real circumstances, total counting rate above approximately one MeV threshold is about tens to one-three hundreds[2] events per second. Therefore, during, for example one month of irradiation about $30*10^5*100$ =3e+08 multi - parameter events are written to the hard disk during a typical SHE experiment at the DGFRS.

From the viewpoint of the detection of rare decays and background suppression, this success was achieved due to the application of a new radical technique – the method of active correlations [8-17]. The method employs search in a real-time mode for a pointer to a probable energy-time-position correlation like recoil-alpha for switching the beam off

---

[1] Depending on the reaction asymmetry ( projectile to target mass ratio)

## The Dubna Gas Filled Recoil Separator (DGFRS)

For the synthesis and study of heavy nuclides, the complete fusion reactions of target nuclei with bombarding projectiles are used. The resulting excited compound nuclei (CN) can deexcite by evaporation of some neutrons, while retaining the total number of protons. Recoil separators are widely used to transport EVRs from the target to the detection system, while simultaneously suppressing the background products of other reaction, incident beam of ions, and scattered target nuclei. A distinctive feature of gas-filled separators [18-27] is the fact that atoms recoiling from the target with the broad distribution of high charge states interact with the gas such that both average charge and dispersion are reduced. The decrease of average charge of EVRs results in their larger rigidity in the magnetic field in comparison with the background ions. Thus, EVRs can be rapidly separated in flight from unwanted reaction products and collected at detection system. From the viewpoint of the separator design D-Q-Q (dipole magnet and two quadrupole lenses) is applied (Fig.1).

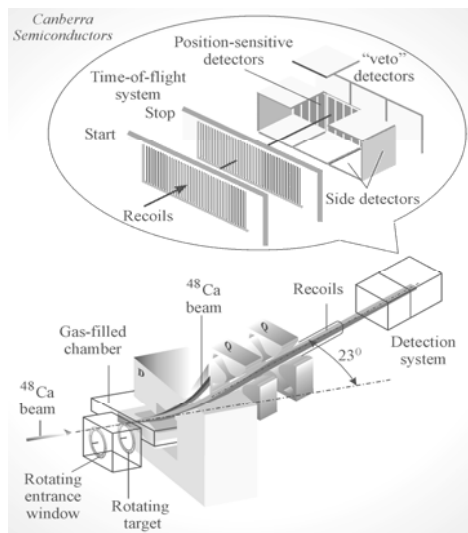

---

[2] Including events having only TOF(time-of-flight) signal and zero energy ( below energy threshold)

**FIGURE.1** The Dubna GasFilled Recoil Separator schematics. The detecting module is shown in the upper part of the figure.

The simple but new idea of the algorithm is aimed at searching in real-time mode of time-energy-position recoil-alpha links, using the discrete representation of the resistive layer of the position sensitive PIPS detector separately for signals like "recoil" and "alpha-particle". So, the real PIPS detector is represented in the PC's RAM in the form of two matrixes, one for the recoils (static) and one for alpha-particles (dynamic). Those elements are filled by values of elapsed times of the given events. The second index number of the matrix element is defined from the vertical position, whereas the first index is in fact strip number (1…12). In each case of "alpha-particle" detection, a comparison with "recoil"-matrix is made, involving neighboring elements (+/-3). If the minimum time is less or equal to the setting time, the system turns on the beam chopper which deflects the heavy ion beam in the injection line of the cyclotron for a 1-5 min. The next step of the PC code ignores the vertical position of the forthcoming alpha-particle during the beam-off interval. If such decay takes place in the same strip that generated the pause, the duration of the beam-off interval is prolonged up to 10-30 min. In the **Fig.2 a,b** schematic of the algorithm and flowchart of the method application are shown.

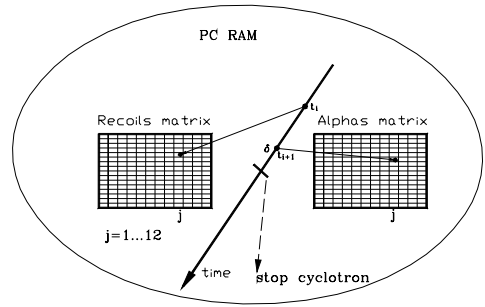

**FIGURE 2a** Schematic of the algorithm. EVR and alpha-particle matrixes are shown.

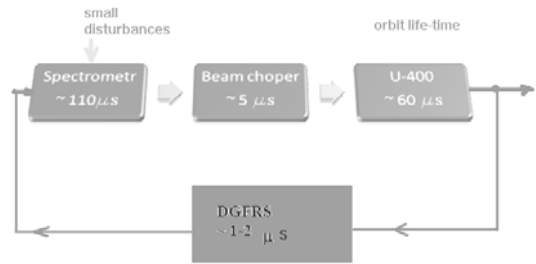

**FIGURE 2b** Flowchart of the beam-interrupting process

## Method Application In Heavy-Ion Induced Nuclear Reactions

Before application in the long-term experiments, the algorithm and technique described above have been tested in $^{48}$Ca + $^{206}$Pb → $^{252}$No+2n and $^{48}$Ca + $^{nat}$Yt → Th$^*$ nuclear reaction tests[3]. During the last 8 years the mentioned method was successfully applied in the HI induced nuclear reactions:

$^{238}$U + $^{48}$Ca → $^{286-x}$112+ xn; $^{242,244}$Pu + $^{48}$Ca → $^{290,292-x}$114 + xn;

$^{245,248}$Cm + $^{48}$Ca → $^{293,296-x}$116+xn; $^{243}$Am + $^{48}$Ca → $^{291-3,4}$115+3,4n ;

$^{237}$Np + $^{48}$Ca → $^{282}$113+3n; $^{249}$Cf + $^{48}$Ca → $^{294}$118+3n;

$^{226}$Ra + $^{48}$Ca → $^{270}$Hs + 4n.

For instance, in the **Fig.3** the results of application is shown for $^{237}$Np+$^{48}$Ca complete fusion reaction and in the **Tab.1** the parameters of radical backgrounds suppression factor are shown in the column 2. The last parameter is defined as a ratio to all alpha particle imitator signal number within the definite energy interval to one, measured in the beam OFF pauses.

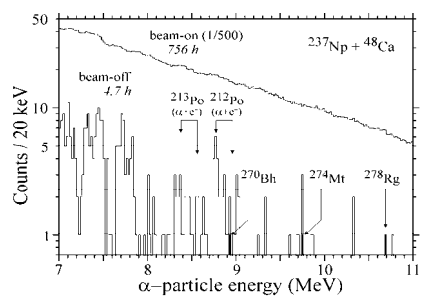

**FIGURE 3** Beam-OFF spectrum of alpha decays were measured in $^{237}$Np+$^{48}$Ca complete fusion reaction .Energy ranges: for EVR (7, 18) and for alpha particles (9.9, 11.4) MeV. Correlation time was 1.5 s for vertical position window 2.8 mm. Beam-Off intervals were of 10 s.

---

[3] Reaction $^{48}$Ca + $^{nat}$Yt → Th$^*$ was used to calibrate alpha-particle energy scale.

**Table 1. Typical suppression factors when "active correlations" technique is applied.**

| Reaction | *An integral suppression factor (9-11 MeV)* | Energy correlation interval (Eα, MeV) | Correlation time, s EVR - α | Beam pause, min |
|---|---|---|---|---|
| $^{238}$U+ $^{48}$Ca→ 112 | *9,5 e+03* | 9,43 – 9,63/ 10,3-11,8 | 12/0,3 | 1 |
| $^{242}$Pu+$^{48}$Ca→114 | *4 e +03* | 9,9 – 10,35 | 4 | 1 |
| $^{245}$Cm+$^{48}$Ca→116 | *1,5 e+04* | 9,9 - 11 | 1 | 1 |
| $^{243}$Am+$^{48}$Ca→115 | *2.0 e+04* | 9,6 - 11 | 8 | 2 |
| $^{249}$Cf+$^{48}$Ca→118 | *1,1 e+04* | 9,9 - 12 | 1 | 1 |

## SUMMARY

A new radical method of "active correlations" is developed, tested and successfully applied in the heavy ion-induced complete fusion nuclear reactions. The latter is based on: - theoretical models, EVR spectra simulations and empirical relations obtained from test reactions, real-time matrix algorithm used to search for pointer to potential forthcoming correlated sequence, DGFRS PC-based detection system, U-400 main FLNR cyclotron complex.

The application of the system has shown that:

- Detection of recoil-alpha correlated sequences in a real-time mode provides a strong suppression of beam-associated backgrounds, when detecting ultra rare alpha decays. It provides clearer event detection and identification in long-term experiments aimed at the synthesis of SHE
- Loss in the overall experiment efficiency is negligible, whereas an additional integral background suppression factor of about $10^4$ in the energy range near 10 MeV has been achieved.